\documentclass[sigconf]{acmart}

\usepackage{multicol}
\usepackage{hyperref}
\usepackage{enumitem}
\usepackage{listings}
\usepackage{subfigure}
\AtBeginDocument{%
  \providecommand\BibTeX{{%
    \normalfont B\kern-0.5em{\scshape i\kern-0.25em b}\kern-0.8em\TeX}}}

\copyrightyear{2021} 
\acmYear{2021} 
\setcopyright{acmcopyright}\acmConference[VINCI '21]{The 14th International Symposium on Visual Information Communication and Interaction}{September 6--8, 2021}{Potsdam, Germany}
\acmBooktitle{The 14th International Symposium on Visual Information Communication and Interaction (VINCI '21), September 6--8, 2021, Potsdam, Germany}
\acmPrice{15.00}
\acmDOI{10.1145/3481549.3481562}
\acmISBN{978-1-4503-8647-0/21/09}

\begin{document}


\title[An Open Source Interactive Visual Analytics Tool for Comparative Programming Comprehension]{An Open Source Interactive Visual Analytics Tool \\for Comparative Programming Comprehension}


\author{Ayush Kumar}
\authornote{Corresponding Author: ayush\_kumar@meei.harvard.edu}
\authornotemark[1]
\affiliation{%
 \institution{Schepens Eye Research Institute,
Harvard Medical School MA}
  \country{USA}
}

\author{Ashish Kumar}
\affiliation{%
 \institution{Stony Brook University, NY}
   \country{USA}
 }

\author{Aakanksha Prasad}
\affiliation{%
 \institution{Stony Brook University, NY}
   \country{USA}
}

\author{Michael Burch}
\affiliation{%
 \institution{University of Applied Sciences of the Grisons, Chur}
   \country{Switzerland}
}

\author{Shenghui Cheng}
\affiliation{%
 \institution{Westlake University, Hangzhou }
   \country{China}
}

\author{Klaus Mueller}
\affiliation{%
\institution{Stony Brook University, NY}
  \country{USA}
}

\renewcommand{\shortauthors}{Kumar et al.}

\begin{abstract}
This paper proposes a visual analytics tool consisting of several views and perspectives on eye movement data collected during code reading tasks when writing computer programs. Hence the focus of this work is on code and program comprehension. The source code is shown as a visual stimulus. It can be inspected in combination with overlaid scanpaths in which the saccades can be visually encoded in several forms, including straight, curved, and orthogonal lines, modifiable by interaction techniques. The tool supports interaction techniques like filter functions, aggregations, data sampling, and many more. We illustrate the usefulness of our tool by applying it to the eye movements of 216 programmers of multiple expertise levels that were collected during two code comprehension tasks. Our tool helped to analyze the difference between the strategic program comprehension of programmers based on their demographic background, time taken to complete the task, choice of programming task, and expertise.
\end{abstract}



\begin{teaserfigure}
  \centering
  \includegraphics[width=0.98\textwidth]{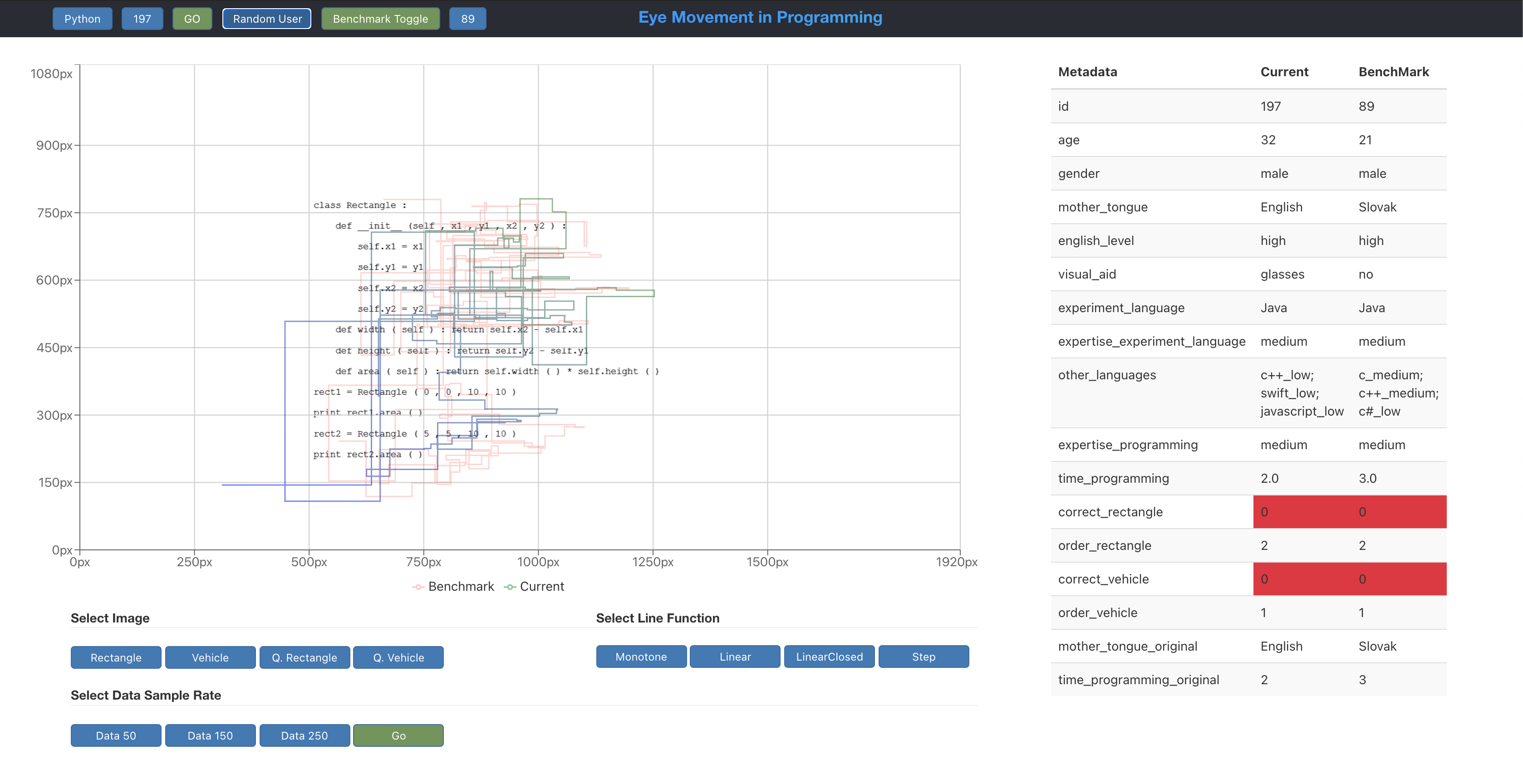}
  \caption{The visual analytics tool with major views and perspectives, for example on the code and the scanpaths, and some of its interactive functionalities. A live demo of the visual analytics tool can be found on  \underline{\href{https://ashvtol.github.io/emip_demo}{\textbf{EMIP Demo}}}}
  \label{fig:teaser}
\end{teaserfigure}

\maketitle

\section{Introduction}

Eye-tracking is a powerful technology to measure and record eye movements over space and time~\cite{Duchowski:03, Holmqvist:11}. Such eye movements can help investigate visual attention strategies~\cite{Andrienko:12,kumar2021eyefix,kumar2020demoeyesac} among the participants to any visual stimulus to find design flaws and, based on them, improve a visual stimulus. This holds not only for image- or video-based stimuli~\cite{Kurzhals:14} but even for text-based stimuli in which eye-tracking participants are confronted by reading tasks~\cite{Oquist:07,jbara2017programmers} or task-based image stimuli where gazes are used to classify the task performed~\cite{kumar2019task, kumar2020challenges}. A particular reading task occurs in software development~\cite{Sharafi:22} in which various lines of source code must be understood to locate bugs that would make the code error-prone and cause semantic and syntactic runtime errors. 

However, it is pretty challenging due to the sheer size, complexity, and spatio-temporal nature of eye movement data~\cite{Burch_ETVABook:22,kumar2020visualthesis} to analyze and visualize the data to detect insights and knowledge in it. This requires a visual analytics tool~\cite{Keim:12} consisting of algorithmic and visual support to explore the data from several dimensions like space, time, and participants, in our case, code developers or programmers. To not let alone the users of such a tool with one view on the data, it is of particular benefit to equip the tool with various visual encodings of the same data, for example, the representation of the scanpaths in either straight, curved, or orthogonal shapes~\cite{Holten:11} incorporating several interaction techniques~\cite{Yi:07} (see Figure~\ref{fig: teaser}). 

In this paper, we introduce a tool that can load eye movement data with a static source code stimulus, transform, filter, and cluster such data, and provide visual perspectives on the data with interaction techniques to navigate the data to detect patterns and anomalies worth investigating, for example, to improve the code based on programmers' visual attention strategies. During the tool’s design, we focused on several significant aspects that partially guided the creation of the tool.

\begin{itemize}[leftmargin=*]
	\setlength{\itemsep}{-2pt}
	\item \textbf{Data:} We support the loading, storing, and transforming of a dataset related to eye movements recorded during programming.
	\item \textbf{Visualization:} We allow the visualization of fixations and saccades for one or several programmers, for example, to understand the visual attention behavior of beginners, experts, and professionals in coding.
	\item \textbf{Algorithms:} To further bring the data into a task-solving and useful form, we incorporated algorithmic approaches to filter, group, and cluster the data.
	\item \textbf{Linking:} As a visual analytics tool consists of a wealth of linked concepts, we also combine visualizations, algorithms, and interactions with the users in the loop to find patterns and anomalies.
\end{itemize}

To illustrate the tool’s usefulness, we apply it to the Eye Movements In Programming (EMIP) dataset~\cite{bednarik2020emip} that contains the eye movements of 216 programmers of multiple expertise levels was collected during two code comprehension tasks. The primary result of this line of research is summarizing the programmers' behavior and strategy analysis. It may be easier to visualize each programmer's flow and gaze attention and compare their strategies in terms of their expertise with our approach than by just relying on the manual inspection of the programmer’s behavior. Such summaries reflect which parts of the code a programmer focused on, for example, the body, arguments, or return types, to mention a few. Although it is difficult to condense these outcomes, one can see regions of specific focus in the case of experienced vs. non-experienced programmers~\cite{Jessup:21}.
\section{Related Work}

Implementing source code is a challenging task in software engineering~\cite{Diehl:07}, but even more, the comprehension~\cite{Bhattacharjee:22} of such textual input can be more challenging, for example, if the code is quite long and complex or if the person inspecting the code does not have an adequate expertise level to solve such tasks~\cite{Jessup:21}. Moreover, code comprehension also depends on aspects like the programming language, the programming or integrated development environment (IDE), and other circumstances like the programmers' state of mind and the places at which they try to solve such code comprehension tasks. A lot of research has been and is going on focusing on code comprehension in the software engineering domain like in maintenance~\cite{Saiyd:17} with several programming paradigms~\cite{9576333}, also including eye tracking~\cite{Rodeghero:15, Rodeghero:19} as a powerful technology to find visual attention patterns during task solving~\cite{sharafi2020practical}.

However, although the recording of eye movement data~\cite{Duchowski:03, Holmqvist:11} in the software engineering domain~\cite{Jessup:21, Rodeghero:15,sharafi2015systematic,sharafi2020practical} provided some interesting research results, the visual support to identify patterns~\cite{Andrienko:12} in such domain-specific eye movement data is quite rare. For example, Shaffer et al.~\cite{shaffer2015itrace} create the iTrace tool to show eye-tracking data in software artifacts, for example, to support software engineers with typical software engineering tasks, to which the programming of code also belongs to~\cite{obaidellah2018survey,bednarik2020emip}. Jessup et al.~\cite{Jessup:21} investigated eye tracking for understanding expert and novice programmers focusing on code comprehension. Although their approach is quite intriguing, they did not include interactive visualizations showing the code stimulus and programmers' scanpaths in several forms. One approach attempts to more or less summarize the findings based on eye movement data~\cite{Rodeghero:14}.

Negatively, visualization or visual analytics tools~\cite{Keim:12} do not exist many for the application field of code comprehension, some of them support some naive visualization techniques, but they do not link algorithmic approaches, visualizations, and interaction techniques with programmers-in-the-loop. One approach was presented by Burch et al.~\cite{Burch_IV:15} who visualized source code to identify similar code fragments of large software systems in one view. The problem is that the visualization is based directly on the code, not including eye movement data.

\section{Dataset}

To show the usefulness of our tool, we use the EMIP dataset~\cite{bednarik2020emip}, released in August 2020. It consists of data from 216 programmers from varying backgrounds, demographics, cultures, and different levels of expertise in programming. Also, this dataset is an example of an international collaborative effort involving eleven researcher teams from eight countries and four continents.

The dataset records the stimuli presented to participants using an SMI laptop with an SMI RED250 mobile eye tracker (sampling rate 250 Hz). The screen resolution was 1920×1080 px. The experiment was built with an SMI Experimental Suite Scientific Premium.

The multiple features of the dataset are:
Time (timestamp of the sample), Type (SAMPLE or MESSAGE), 
L Raw X [px]: horizontal pupil position – left eye,
L Raw Y [px]: vertical pupil position – left eye,
R Raw X [px]: horizontal pupil position – right eye,
R Raw Y [px]: vertical pupil position – right eye,
L POR X [px]: point-of regard in the X axis in pixels – left eye,
L POR Y [px]: point-of regard in the Y axis in pixels – left eye,
R POR X [px]: point-of regard in the X axis in pixels – right eye,
R POR Y [px]: point-of regard in the Y axis in pixels – right eye,

The dataset contains a tsv-file with raw eye movement data for each participant. The file header provides information like the version of the IDF Converter used, the number of samples, and the coordinates of the calibration points. 

\begin{figure*}
    \centering
    \includegraphics[width=0.9\textwidth]{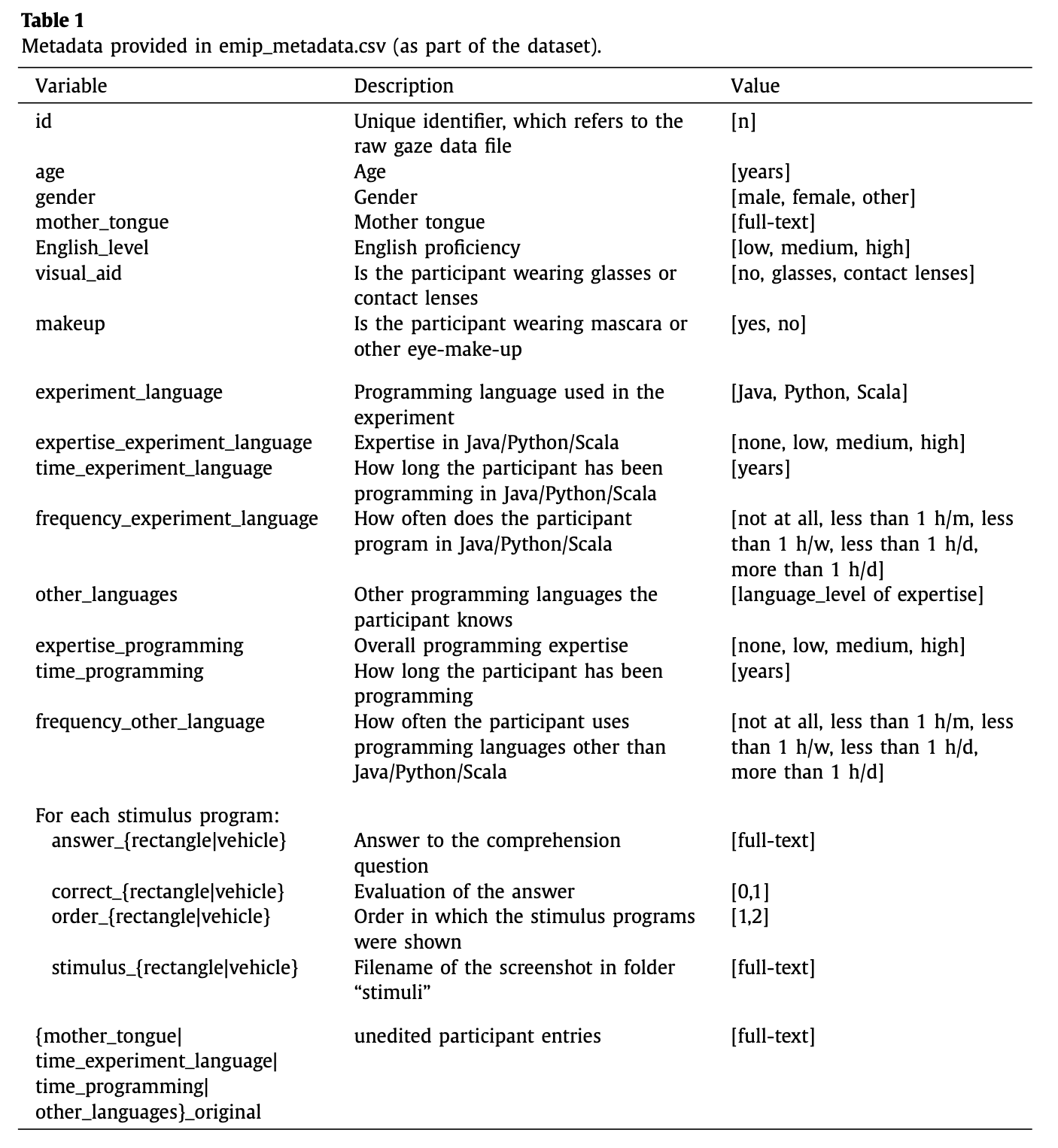}
    \caption{Features in the metadata~\cite{bednarik2020emip}}
    \label{fig:t1}
\end{figure*}

The metadata features are consolidated in Figure~\ref{fig:t1} to understand the data quickly.

\section{Scope of Visual Analytic}
With the increasing interest in eye-movement tracking, there is a also need to consider its impact and assess various use cases by merging hypotheses and results. One approach to achieve this is arranging an enormous, freely accessible dataset that can be used to check existing hypotheses and form new ones.

\begin{figure}
    \centering
    \includegraphics[width=0.5\textwidth]{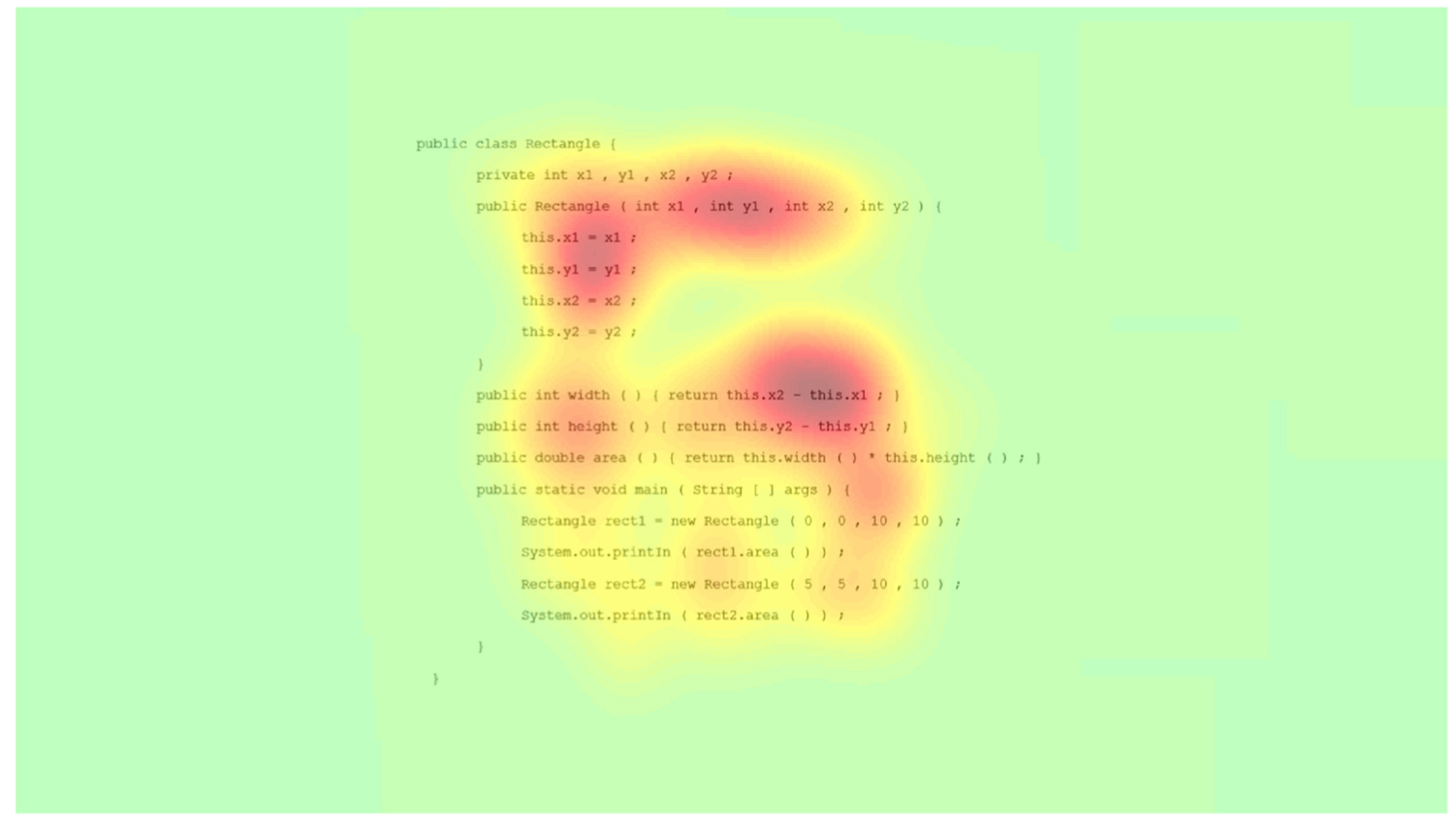}
    \caption{This plot shows gaze density maps of a participant who was looking at a piece of code.}
    \label{fig:tVa}
\end{figure}

That said, visual analytics plays an essential role in this motivation. For example, we can use the visualization plots as shown in Figure~\ref{fig:tVa}, given the eye-movement data that emphasizes the key area of visual attention.

The red regions in Figure~\ref{fig:tVa} are the high density of fixations, the green is low, and the other colors correspond to the fixation density between high and low.

These insights are significant in interpreting decisions or understanding the person's attention or thought process. It is deeply entwined with the field of psychology or human behavior and can assist in finding interesting patterns.

\subsection{Expectation}
In this paper, we propose a Visual Analytic tool~\footnote{Open Source Code repository of the VA tool on \underline{\href{https://github.com/ashvtol/EMIP/tree/master}{\textbf{github}}} } that will use the required data from the given dataset repository~\cite{bednarik2020emip}, visualize and visually interact with it to be able to compare the differences in fixation and scanpaths between various groups (beginner, intermediate and experienced, etc.) of participants. We will also compare the variance of the fixations between the different groups to better understand the results' outcome.

\subsection{Data Loading \& Pre-processing Details}
\begin{figure}
    \centering
    \includegraphics[width=9cm]{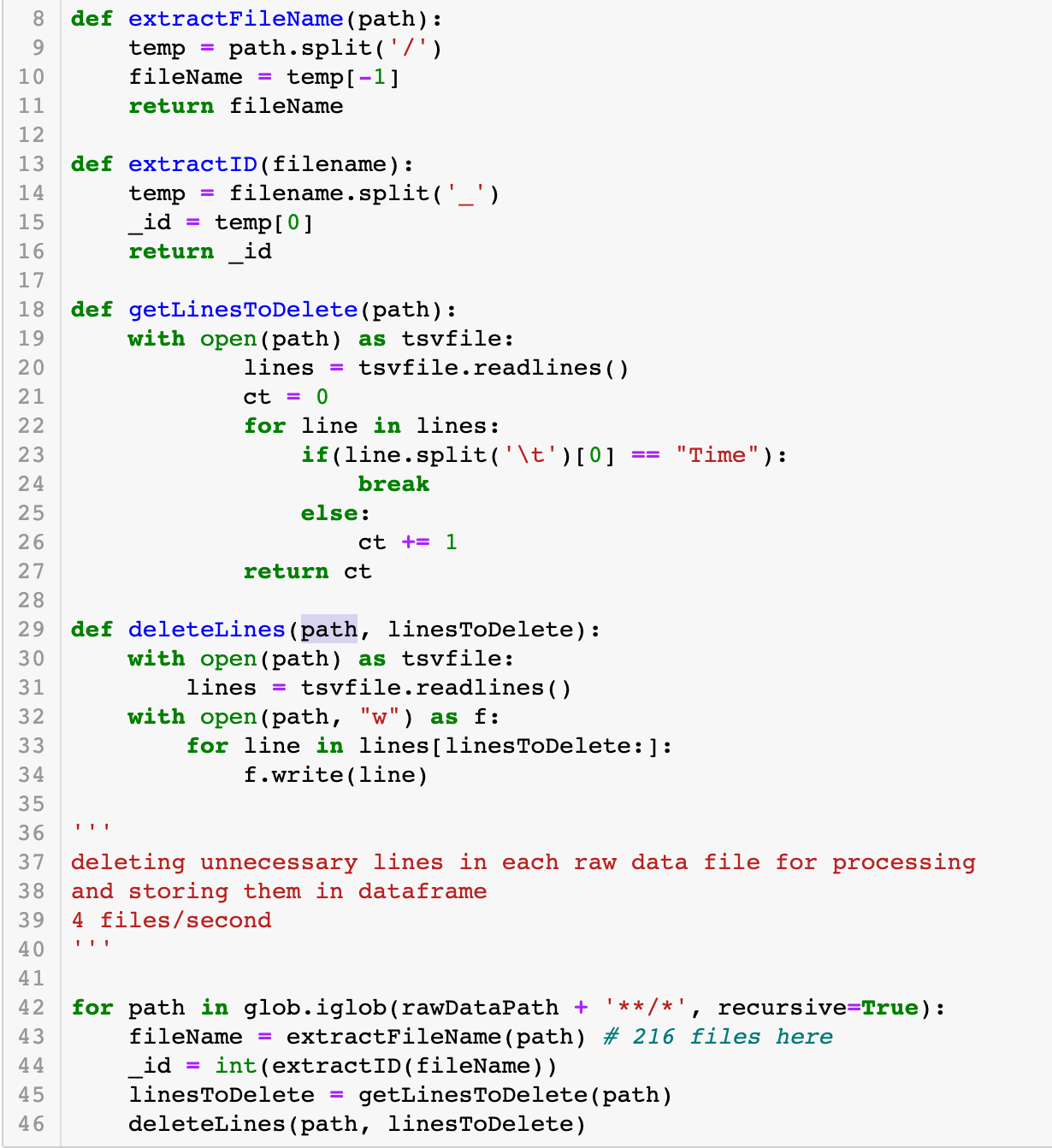}
    \caption{Functions to extract filename, ID, and lines to delete}
    \label{fig:code_sn1}
\end{figure}

To begin with, data~\cite{bednarik2020emip} from 216 samples were loaded using the glob library in Python to pre-process the data. 
Followed by, filtering the required data, i.e., the samples, were extracted from each file by deleting the unnecessary information. The extracted filtered data was then stored in a data frame. Entire code written for this tool is in modular format for maintainability as the code snippet of the same can be shown in Figure~\ref{fig:code_sn1}.

\textbf{Glob.iglob} recursively looks for all the files inside the dataset folder.
First the path of a \textit{TSV} file (Tab Separated File) is passed to the function \textit{extractFileName}, which extracts the name of the file from the given path. Each file name contains the id of the subject participating, so we use\textit{extractID} from the filename for information while fetching the metadata. The function \textit{getLinesToDelete} open each file and keeps storing the range of the lines with unnecessary metadata information to delete, which we won’t be using for visualization. This range is passed to the function \textit{deleteLines} that deletes it and makes these files parsable to the panda's parser.

Further pre-processing was done to extract only the samples of duration when the participant was observing the code snippet of vehicle and rectangle, albeit the language of the code snippet could be \textit{java/scala} from each processed file. To do this, the index of the start and end of the sample of both the image was retrieved when the user was observing those code snippets, and this is done in the function~\textit{extractRevelantLines}.

\begin{figure}
    \centering
    \includegraphics[width=9cm]{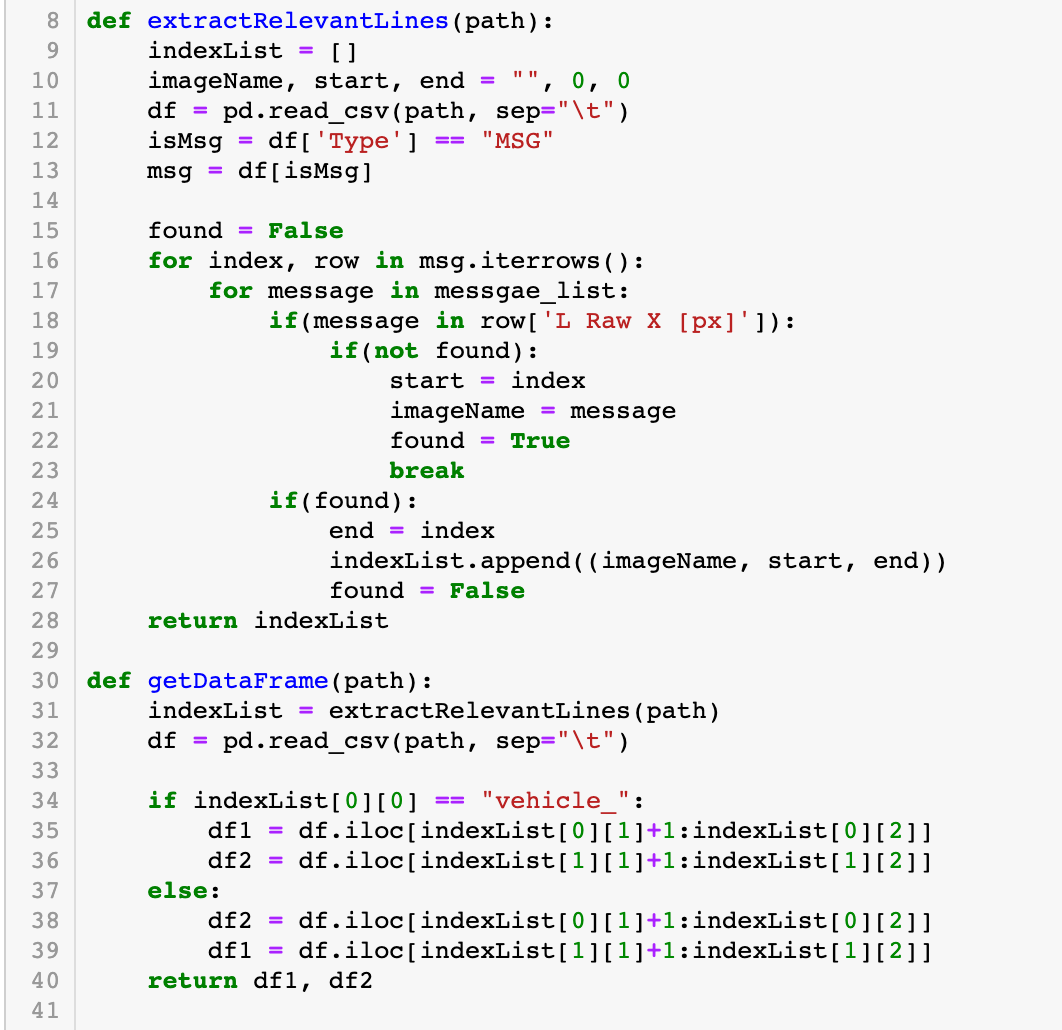}
    \caption{Functions to extract relevant lines and make data-frames}
    \label{ts11}
\end{figure}

Using the range of indexes from \textit{extractRevelantLines}, we create separate data frames for the vehicle and rectangle samples for participants.

\begin{figure}
    \centering
    \includegraphics[width=9cm]{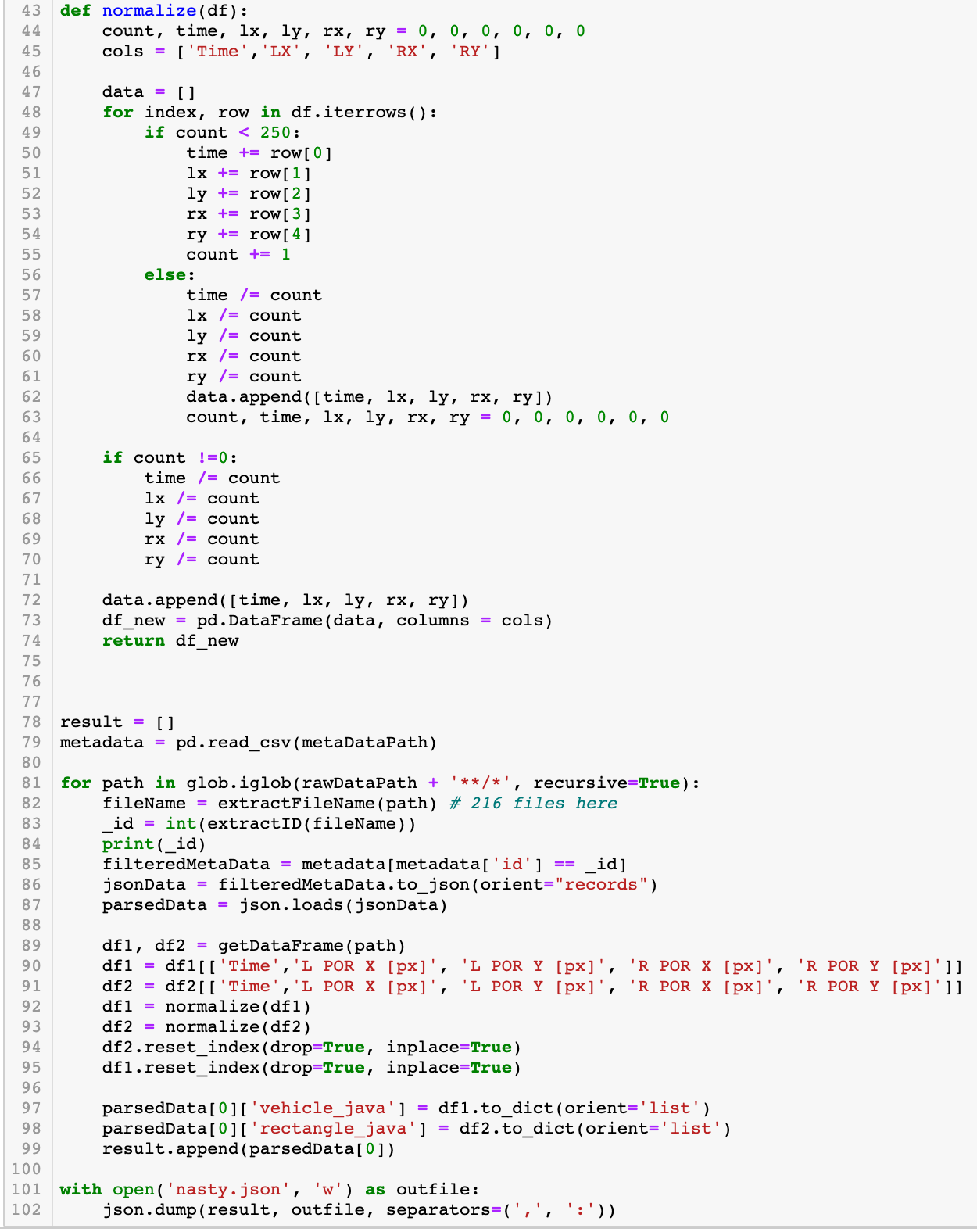}
    \caption{Functions to normalize the data.}
    \label{ts12}
\end{figure}

\begin{figure}
    \centering
    \includegraphics[width=9cm]{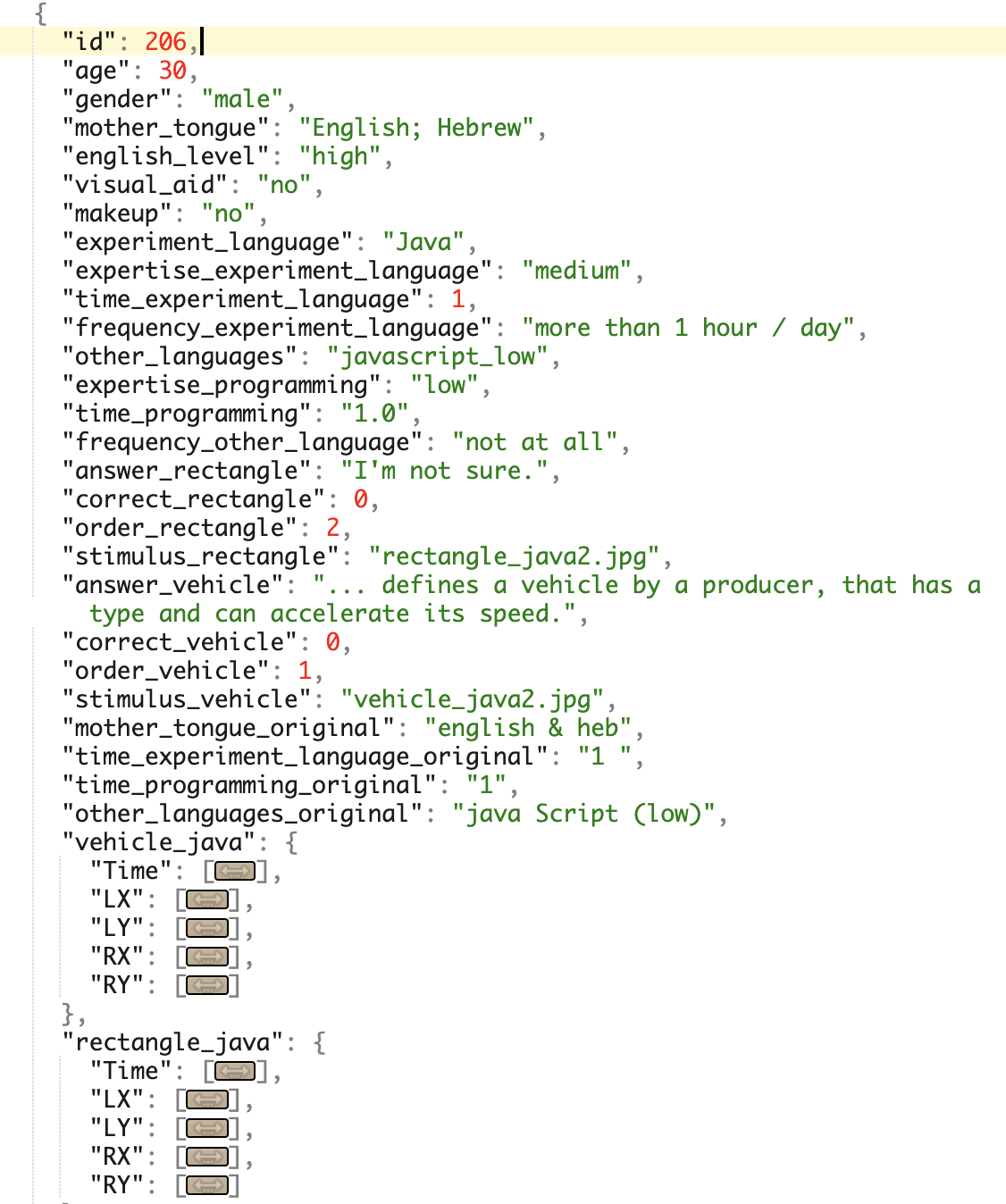}
    \caption{Raw data converted to JSON.}
    \label{ts13}
\end{figure}

Then, we pick the useful features from the file such as columns time, LX [px], LY [px], RX [px], and RY [px], which are normalized using the average of 250 data points, since the frequency of recorded data is \textit{250 Hz}. Normalization is done for each column, and a new data frame is generated from them. Using the normalized data, we then transformed it into a JSON format. Normalization (Figure~\ref{ts12}) occurs in the function \textit{normalize} which returns the new condensed data frame. The data frame of the vehicle and code snippet, along with the metadata, is stored in the result list, which is then later encoded to JSON(Figure~\ref{ts13}), so that it can be easily used in our visual analytic tool.

The condensed JSON encoded data is now reduced to 2.3 MB in size from the initial raw data, which was over 2 Gigabytes. Thus it would make the web tool extra scalable, responsive and faster.

Since the data is recorded at 250 Hz, which means 250 data points are recorded every second, we sample the data as we don't need all the data since most of the data points are identical, and processing all the data points together would be algorithmically challenging in many cases especially eye tracking study with longer duration. We sample the data at regular intervals, say 5 times each second, and the average of these 5 points, and store it as a single point. User can use their own sampling frequency while processing the data using our open source tool. In our case, this incredibly reduces the computation while pre-processing. We have sampled the data 5 times a second, 2 times a second, and one time a second for the demo purpose. Using the three sample rates, we have created three datasets with the advantage of looking at more refined detail and faster pre-processing.

\subsection{Visualization Tool}

The VA tool comprises three main components.
\begin{itemize}[leftmargin=*]
	\setlength{\itemsep}{-2pt}
	\item  Header Menu used to navigate the participant's data.
	\item  Cards that shows the metadata.
	\item  Visualization plot where the gaze is overlaid on top of the images of vehicle and rectangle shown to the participants.
\end{itemize}

\begin{figure}
    \centering
    \includegraphics[width=9cm]{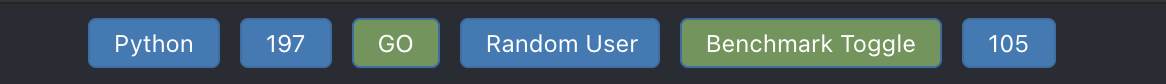}
    \caption{Options to choose language, select and toggle the benchmark participant kid or a random participant in the VA tool.}
    \label{v1}
\end{figure}

Our interactive visual dashboard displays the metadata and gaze plot in pixels. These gaze data points represent the attention points and the eye movement within the image. Using these points, lines were drawn to connect them to find the eye’s movement scanpath. The user can view details regarding a particular participant by selecting the respective IDs and the language of the code from the \textit{dropdown} at the top as shown in Figure~\ref{v1}. The type of experiment can be chosen from the buttons, as shown in the image below. The data of a random participant can also be populated/visualized by clicking on the ‘random user’ button. Also, for a better comparison, the user can select a benchmark participant and toggle the benchmark data on the plot alongside the current participant data (Figure~\ref{v1}).

\begin{figure}
    \centering
    \includegraphics[width=6cm]{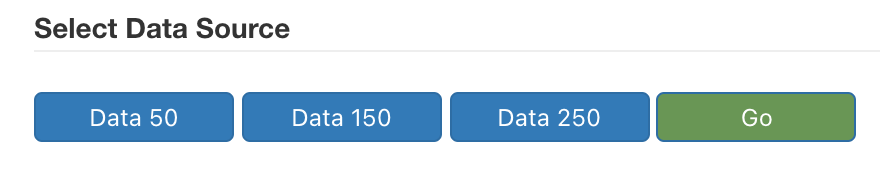}
    \caption{Options to choose samples data points in the VA tool.}
    \label{fig:v12}
\end{figure}

There are more features provided to enhance the user experience and understand the participant’s psyche. The user can choose the rate at which the data is sampled from the options as follows and shown in figure~\ref{fig:v12}:

\begin{itemize}[leftmargin=*]
	\setlength{\itemsep}{-2pt}
	\item  50 (5 times a second),
	\item  150 (2 times a second) and,
	\item  250 (once each second).
\end{itemize}

\begin{figure*}
 $\begin{array}{ccc}
    {\includegraphics[width=0.33\textwidth]{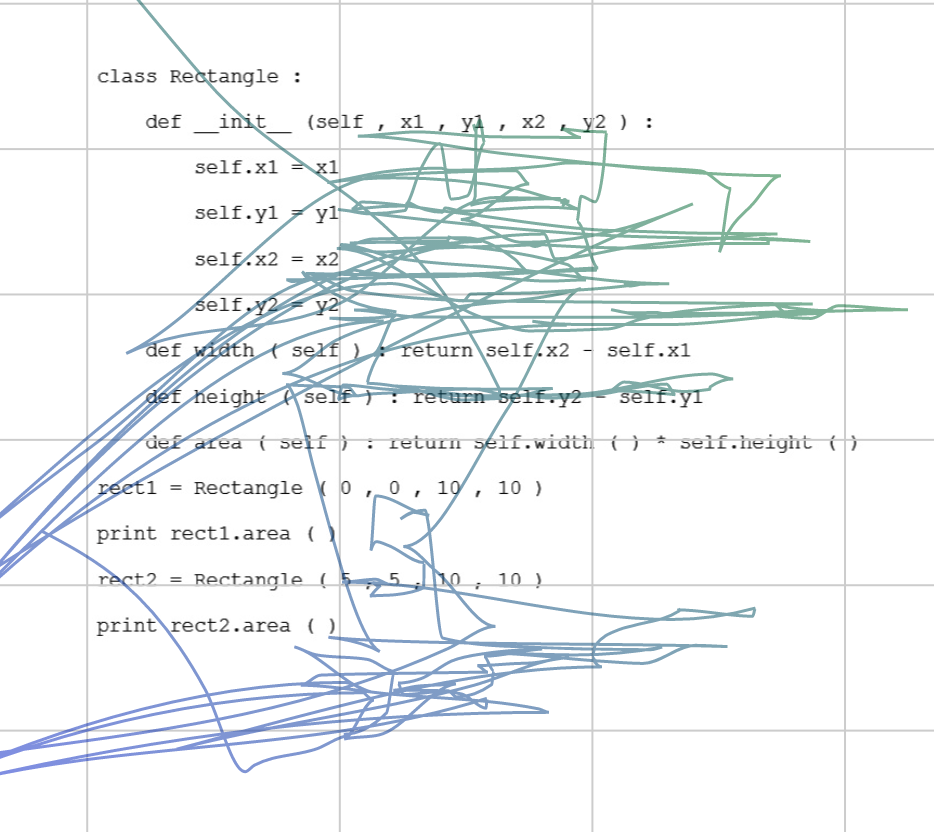}} &
   {\includegraphics[width=0.33\textwidth]{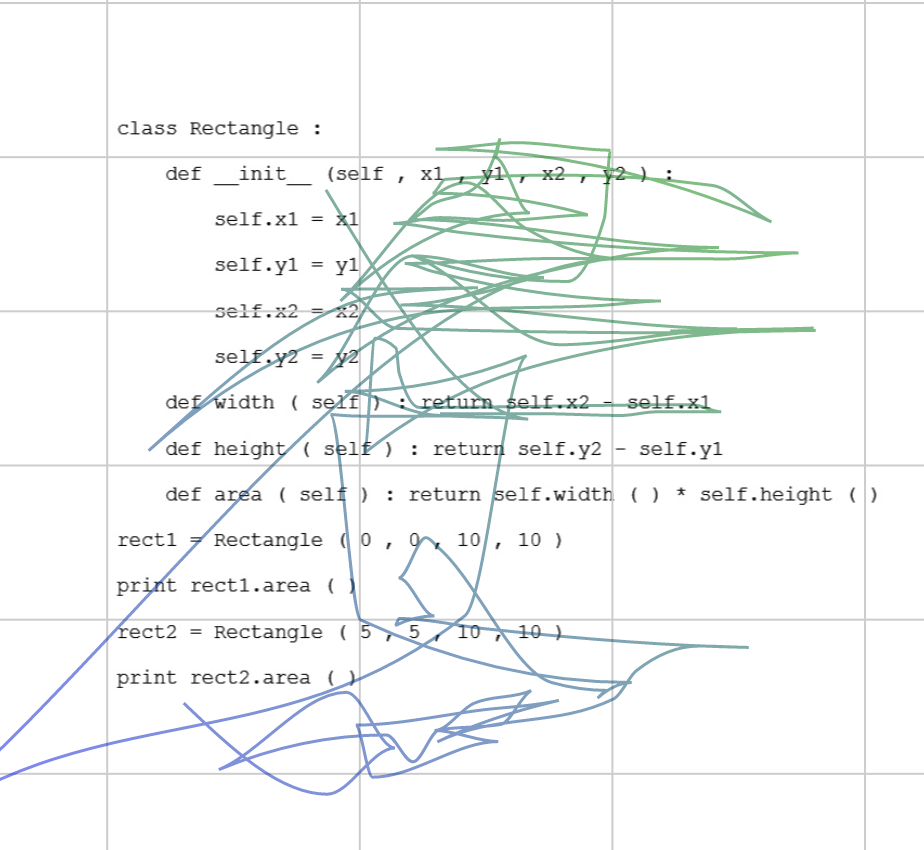}} &
    {\includegraphics[width=0.33\textwidth]{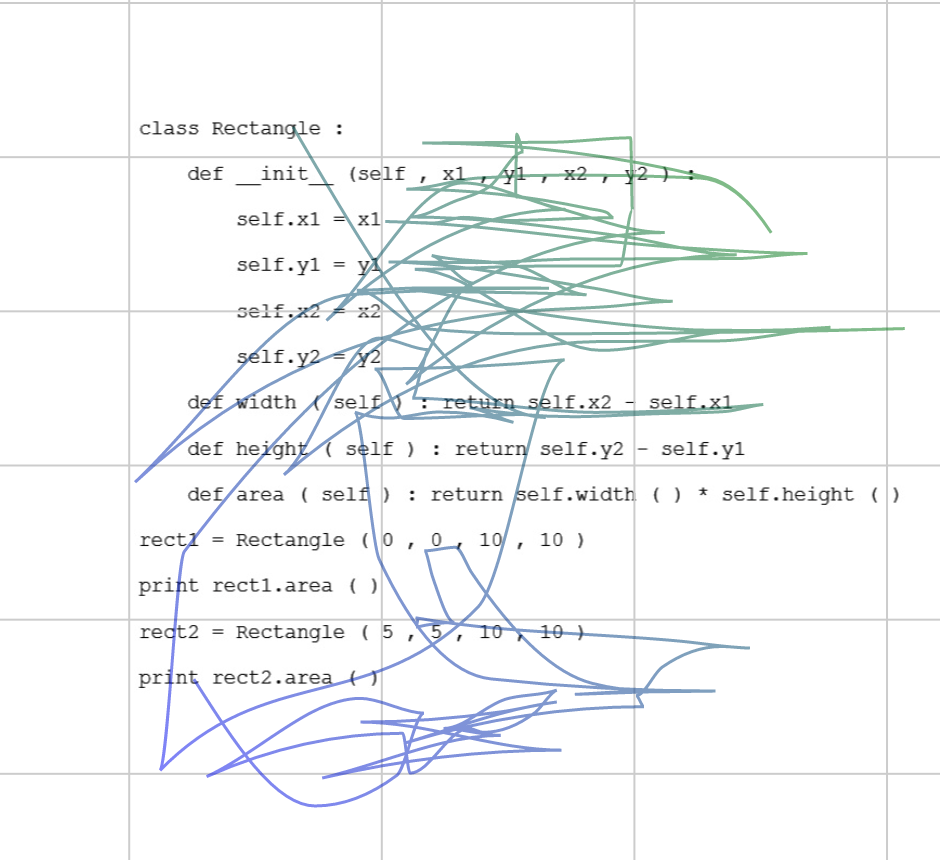}} \\
     \mbox{(a)} & \mbox{(b)} & \mbox{(b)}
 \end{array}$
    \caption{Data Source - with (a) 50 sample points (b) 150 sample points, and (c) 250 sample points}
    \label{fig:sampledata}
\end{figure*}

Choosing the first option helps visualize the participant's area of focus in the image in finer detail due to the presence of more data points. When choosing the third option, the path of eye movement can be envisioned without much clutter.
We experimented with the above three options and decided to provide the user the option to choose the sampling rate. This would help deepen understanding or aid those looking for more detailed information as shown in Figure~\ref{fig:sampledata}.

\begin{figure}
    \centering
    \includegraphics[width=9cm]{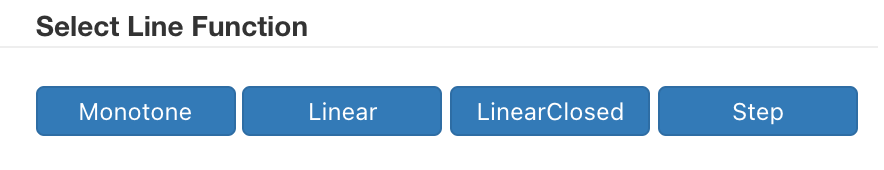}
    \caption{Options to choose line function in the VA tool.}
    \label{v13}
\end{figure}

\begin{figure*}
 $\begin{array}{ccc}
    {\includegraphics[width=0.33\textwidth]{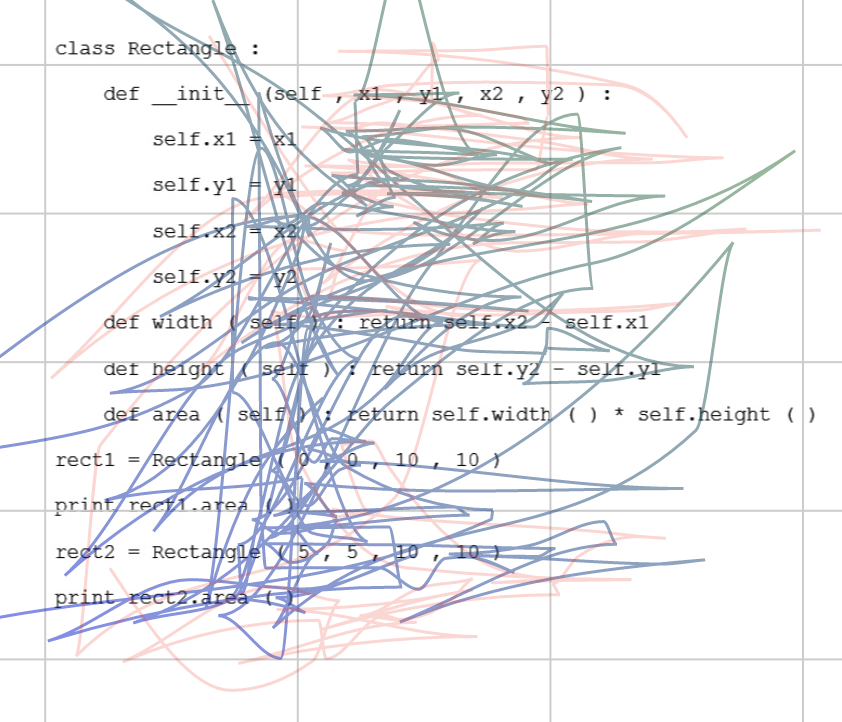}} &
   {\includegraphics[width=0.33\textwidth]{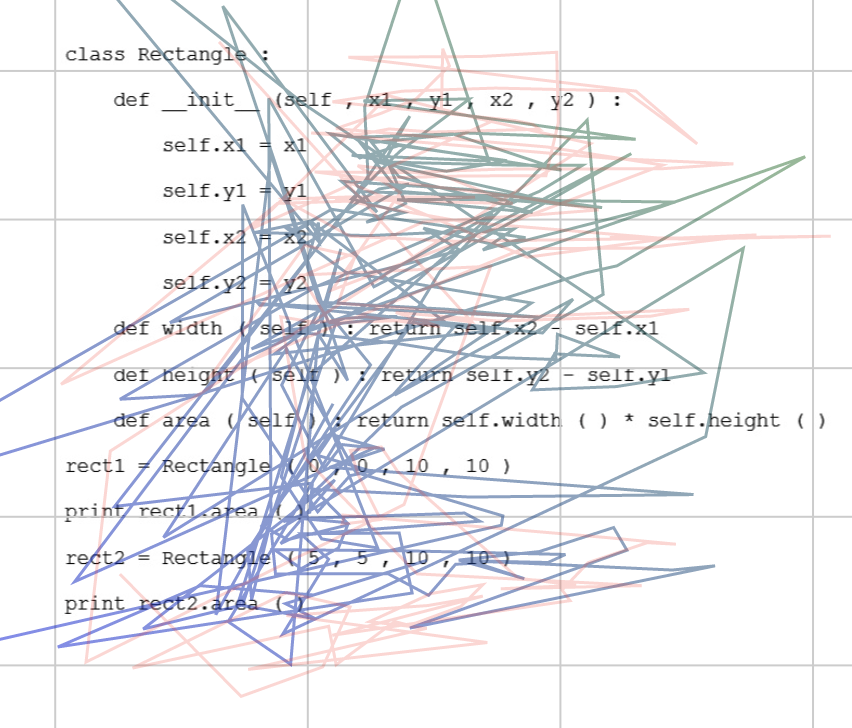}} &
    {\includegraphics[width=0.33\textwidth]{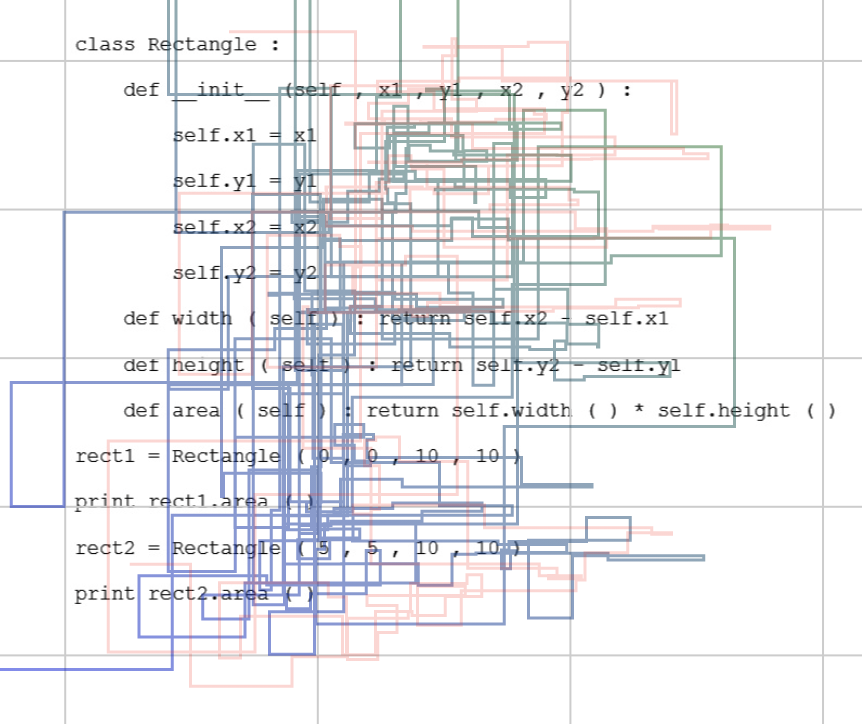}} \\
     \mbox{(a)} & \mbox{(b)} & \mbox{(b)}
 \end{array}$
 \caption{Gaze data visualization using (a) monotone (b) linear, and (c) step function}
 \label{fig:linefunction}
\end{figure*}

Alternatively, one can use choices in line function such as the step function, linear and linear closed, and monotone to change the function as required as shown in Figure~\ref{fig:linefunction}. The above functions interpolate the line between the points (POR) of the eye.

\subsection{Implementation Details}

The visualization tool is developed using React. The flexibility of use and features like virtual-DOM makes it a compelling option. The tool has three main components, header, cards (metadata), and plot. The props selected from the header are passed to the parent and then to the plot. The plot has two components, the Cartesian coordinate and the buttons, which perform mutation on the data and reflect them in the Cartesian coordinates. The plot is made in Recharts.org, a library in react build on top of D3, making it easier to use all the D3 functionalities like line and SVG component rendering.

\begin{figure}
    \centering
    \includegraphics[width=6cm]{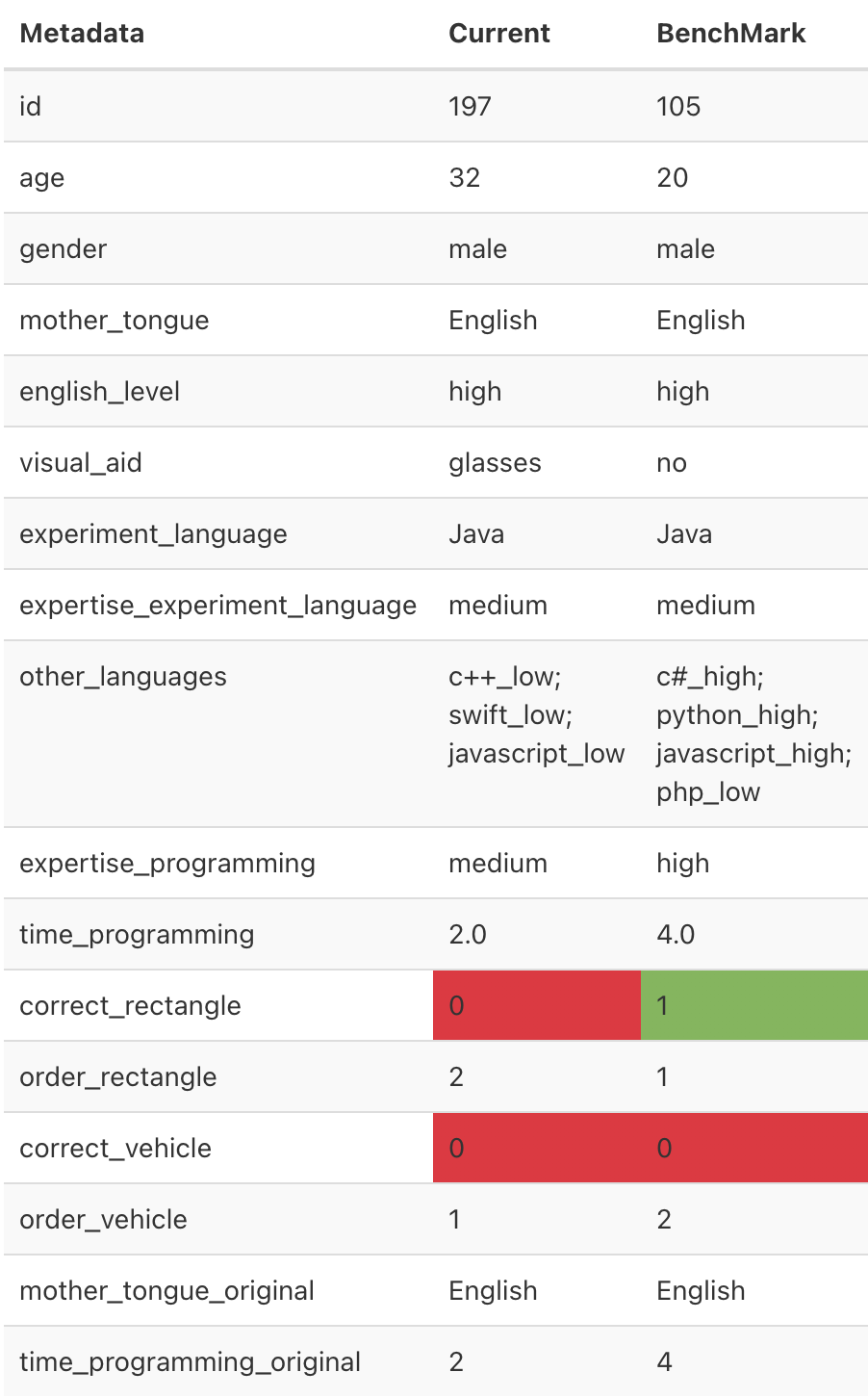}
    \caption{Metadata visualized in the tool for deeper insight during analysis.}
    \label{v16}
\end{figure}

\subsection{Metadata}
The cards on the right of the dashboard depict the information or metadata related to the experiment and the participant. It displays details such as the id, age, gender, level of understanding of English, visual aid (i.e., whether the participant was wearing glasses during the experiment), makeup (presence of any eye makeup), the language of both experiments, and overall programming expertise, the order of the experiment, time spent programming generally and in a specific language, the response to the question after each experiment and the result.

\textbf{Analysis}: This tool helps the user to visualize the flow and concentration of gaze for each participant programmer. Like which parts of the code the user focused on, body, arguments, or return types. Although it is difficult to materialize these outcomes, one can see regions of focus in the case of experienced vs. non-experienced programmers. Also, most participants correctly answered the question in the Rectangle code snippet from the experiment's point of view. But fewer responded to the question about the Vehicle code-snippet correctly regardless of their expertise and eye-movement data.

\begin{figure}
    \centering
    \includegraphics[width=0.49\textwidth]{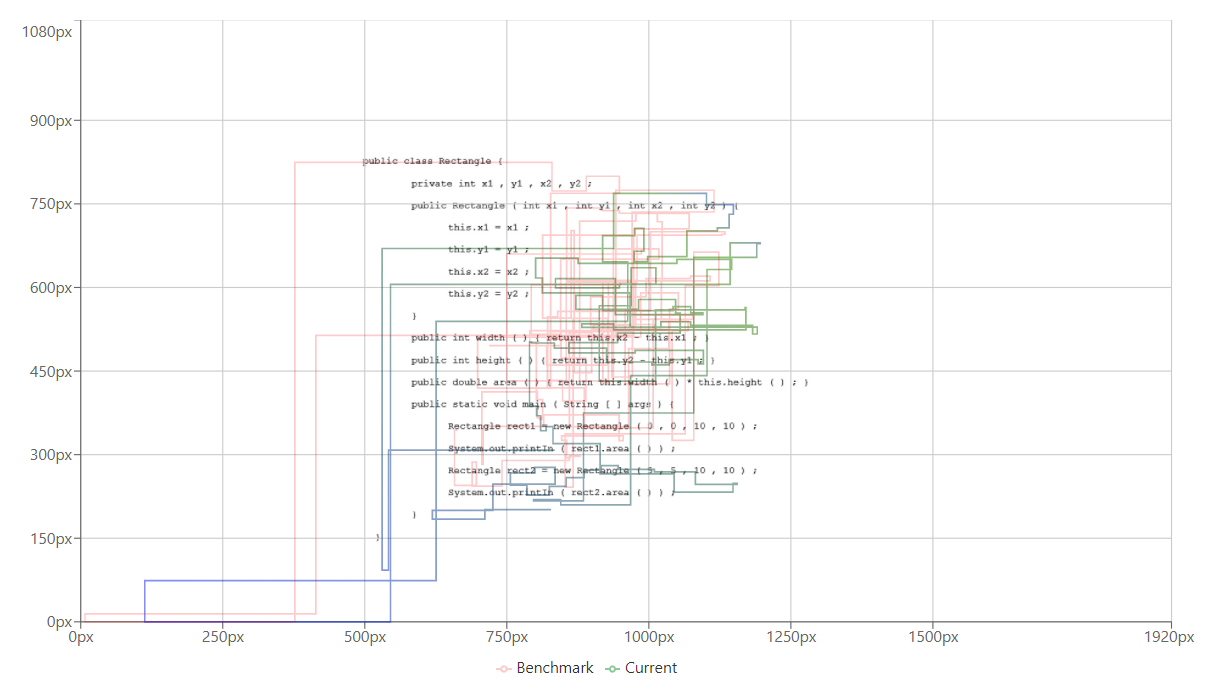}
    \caption{Comparing benchmark participant with the current participant on rectangle code-snippet.}
    \label{v14}
\end{figure}

To analyze the eye movement of a participant, the user can compare the current participant’s eye movement with the benchmark participant’s eye gaze. The benchmark can be either a participant who answered the question correctly and has a high experience or someone who has low experience or responded to the question correctly. The comparison will be instrumental in understanding whether the current participant missed out on going through certain parts of the code that could have answered the question correctly or if participants focused on the unnecessary parts of the code. For example, in the Figure~\ref{v14} (Rectangle, Java), the pink line is the benchmark participant’s gaze, and the blue/green is that of the current participant. The benchmark participant scored the correct answer but not the current participant. We can see that the benchmark participant thoroughly went through the code and focussed their gaze on the constructor’s arguments, the body of the member functions, and the object instantiation. The current participant, however, has only looked at the body of the member function ‘width,’ the second object instantiation, and glanced through the rest of the code. If the current participant had focused on the body of all member functions could have resulted in picking the correct answer for the respective code.
Similarly, one can see that the participants who did not answer correctly did not concentrate on the member function's body but skimmed through the code.

\begin{figure}
    \centering
    \includegraphics[width=0.5\textwidth]{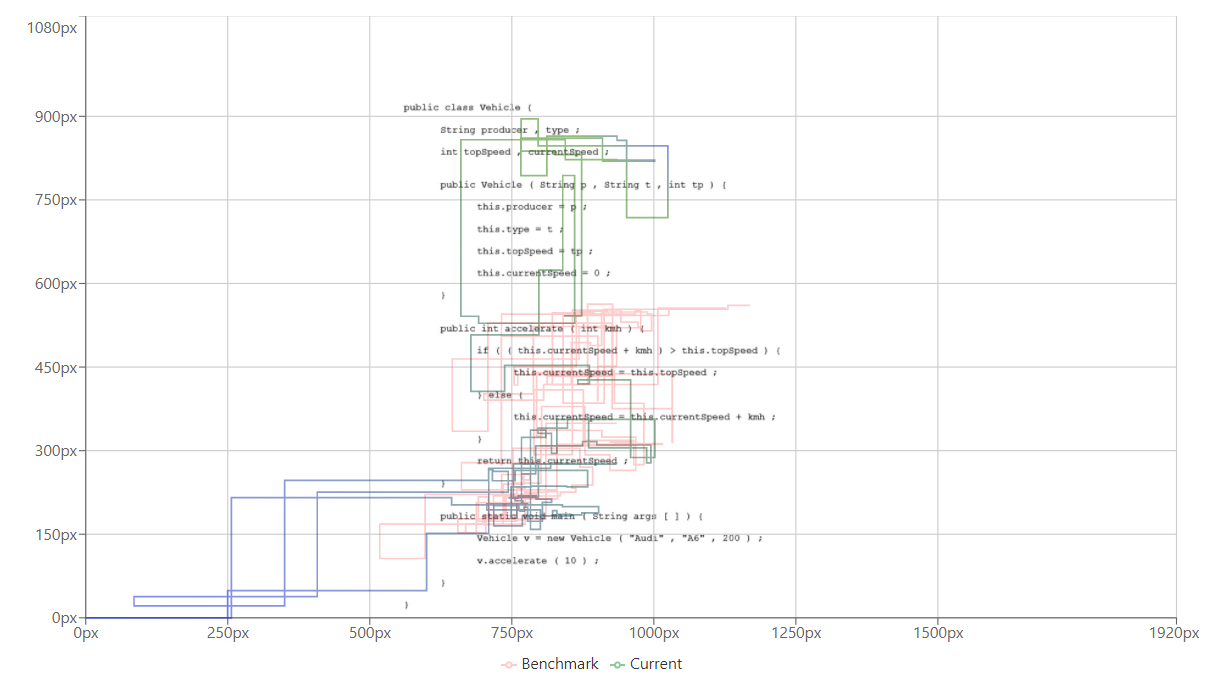}
    \caption{Comparing benchmark participant with the current participant on vehicle code-snippet.}
    \label{v15}
\end{figure}

For experiments that used the Vehicle code as shown in Figure~\ref{v15}, a similar comparison with the benchmark can be made. The figure shows that the current participant spent more time on the data members and the primary method. In contrast, the benchmark participant spent considerable time on the body of the member function and its arguments.

\section{Discussion and Future Work}

We explored the EMIP dataset to validate our open source tool with a demo including preprocessing the raw data, and visualizing it along with the metadata. The tool enable users to select a particular participant id (out of 216 participants). Also, the participants can be selected based on the choice of programming language. The tool also has an option to observe a random participant's data. You can also cycle through the images of the code-snippet shown to the participant, i.e., either vehicle or rectangle. The tool can also see the questions about the code snippets.
To analyze the gaze, you can choose different line functions such as linear, monotone, and steps in the tool to interpolate the data points provided in the dataset. The tool allows changing the data source to different sampling levels to get a finer or less cluttered view of the gaze. To let everyone use our tool, we have also open sourced our code to be used on another dataset including explanation of all the code snippets for respective functionalities.

 Also, source code may inspire various types of low-level eye movement boundaries contrasted with inspecting pictures or understanding exposition, given that code isn't ordinarily perused consecutively and will probably involve repeated visits to especially significant regions. And from the experiment’s point of view, most of the participants correctly answered the question in the Rectangle code snippet. Still, fewer answered the question about the Vehicle code-snippet correctly. Our tool is still in developmental phase and we are planning to make it more user friendly as well as more interactive functionalities to include in it.


\bibliographystyle{ACM-Reference-Format}
\bibliography{sample-base}


\end{document}